\documentclass[twocolumn,aps,prl,showpacs]{revtex4-1}

\usepackage{latexsym,amssymb,float}
\usepackage{setspace}
\usepackage{graphicx}
\usepackage{epsfig}
\usepackage{amsmath}
\usepackage{bm}
\usepackage{appendix}

\begin{document}

\title{Current Eigenmodes and Dephasing in Nanoscopic Quantum Networks}

\author{Tankut Can$^{1}$, Hui Dai$^{1}$, and Dirk K. Morr$^{1,2}$}
\affiliation{$^{1}$ Department of Physics and James Franck
Institute, University of Chicago, Chicago, IL 60637, USA \\
$^{2}$University of Illinois at Chicago, Chicago, IL 60607, USA}
\date{\today}

\begin{abstract}
Using the non-equilibrium Keldysh Green's function formalism, we show that the non-equilibrium charge transport in
nanoscopic quantum networks takes place via {\it current eigenmodes} that possess characteristic spatial patterns. We identify the microscopic relation between the current patterns and the network's electronic structure and topology and demonstrate that these patterns can be selected via gating or constrictions, providing new venues for manipulating charge transport at the nanoscale. Finally, decreasing the dephasing time leads to a smooth evolution of the current patterns from those of a ballistic quantum network to those of a classical resistor network.
\end{abstract}


\maketitle

Understanding charge transport in nanoscale systems \cite{Shi00,Pec04,CLoops,Tod99} has attracted significant interest over the last few
years in the context of molecular electronics \cite{Sol10} and through
recent advances in the fabrication of artificial quantum structures \cite{Top00,Pec03,Are07,Mar07,Man00,Sin11}. The coherent nature of such nanoscopic
systems is of particular importance: it leads to the formation of eigenmodes in the density of states, which are the basis for the engineering
of novel quantum phenomena, such as quantum imaging \cite{Man00}.  The question naturally arises of how the interplay between
the spatial structure of such eigenmodes, dephasing, and the leads' location and width affects the
charge transport through nanostructures.

In this article, we address this question using the non-equilibrium Keldysh
Green's function formalism \cite{theory,Car71a}. We show that nanoscopic quantum networks possess
current eigenmodes exhibiting distinct spatial current patterns that can be selected via gating or through constrictions.
We identify the microscopic relation between these current patterns, the electronic structure (and geometry) of the quantum network, and the leads' location and width for different network topologies.
Moreover, we show that the existence of circulating current loops in these eigenmodes can be described via their enstrophy, and that modes with a large enstrophy in general carry a small total current. Finally, we show that with decreasing dephasing time, the current patterns evolve smoothly from those of a ballistic
quantum network to those of classical resistor network. These results provide important insight into the question of how
charge transport can be controlled and manipulated at the nanoscale.

We consider a two-dimensional quantum network consisting of $N=N_x \times N_y$ sites with Hamiltonian
\begin{eqnarray}
{\cal H}_c & = &  -  t \sum_{{\bf r},{\bf
r}^\prime,\sigma} \; c^\dagger_{{\bf r},\sigma} c_{{\bf r}^\prime,\sigma} - t_{l} \sum_{{\bf l},{\bf l}^{\prime},\sigma} \; d^\dagger_{{\bf l},\sigma} d_{{\bf l}', \sigma}  \nonumber \\
 & & - t_h \sum_{{\bf r},{\bf l},\sigma} \; \left( c^\dagger_{{\bf r},\sigma} d_{{\bf l}, \sigma} + d^\dagger_{{\bf l},\sigma} c_{{\bf r}, \sigma} \right) \nonumber \\
& & + \sum_{\bf r} \omega_0 a_{\bf r}^\dagger a_{\bf r} + g \sum_{\bf r, \sigma} (a_{\bf r}^\dagger + a_{\bf r}) c^\dagger_{{\bf r},\sigma} c_{{\bf r},\sigma}
\label{eq:Hamiltonian}
\end{eqnarray}
Here, $c^\dagger_{{\bf r},\sigma}$ ($d^\dagger_{{\bf l}, \sigma}$)
creates an electron with spin $\sigma$ at site
{\bf r} in the network ({\bf l} in the leads). $t$, $t_{l}$, and $t_h$ are the hopping matrix elements
between neighboring sites in the network, in the leads, and between the network and the leads, respectively. Below, we set for
concreteness $t_{l}/t=10$, and $t_h/t=0.1$. The last term describes the interaction of electrons with a local phonon mode with energy $\omega_0$.

The current between adjacent sites ${\bf r}$,${\bf
r}^\prime$ in the network is induced by different
chemical potentials, $\mu_{L,R}$ in the left and right
leads, and given by \cite{Car71a}
\begin{equation}
I_{\bf r  r^\prime}=-2 \frac{e}{\hbar} \; t
\intop_{-\infty}^{+\infty}\frac{d\omega}{2\pi}{\rm Re} \left[{\hat
G}^K_{\bf r  r^\prime}(\omega)\right] \ . \label{eq:Current}
\end{equation}
Here, ${\hat G}^{K}$ is the full Keldysh
Green's function matrix of the network and leads, which
accounts for the electronic
hopping, the leads' electronic structure, and the electron-phonon interaction.
%
%
\begin{figure*} \begin{center}
\includegraphics[width=17cm,clip]
{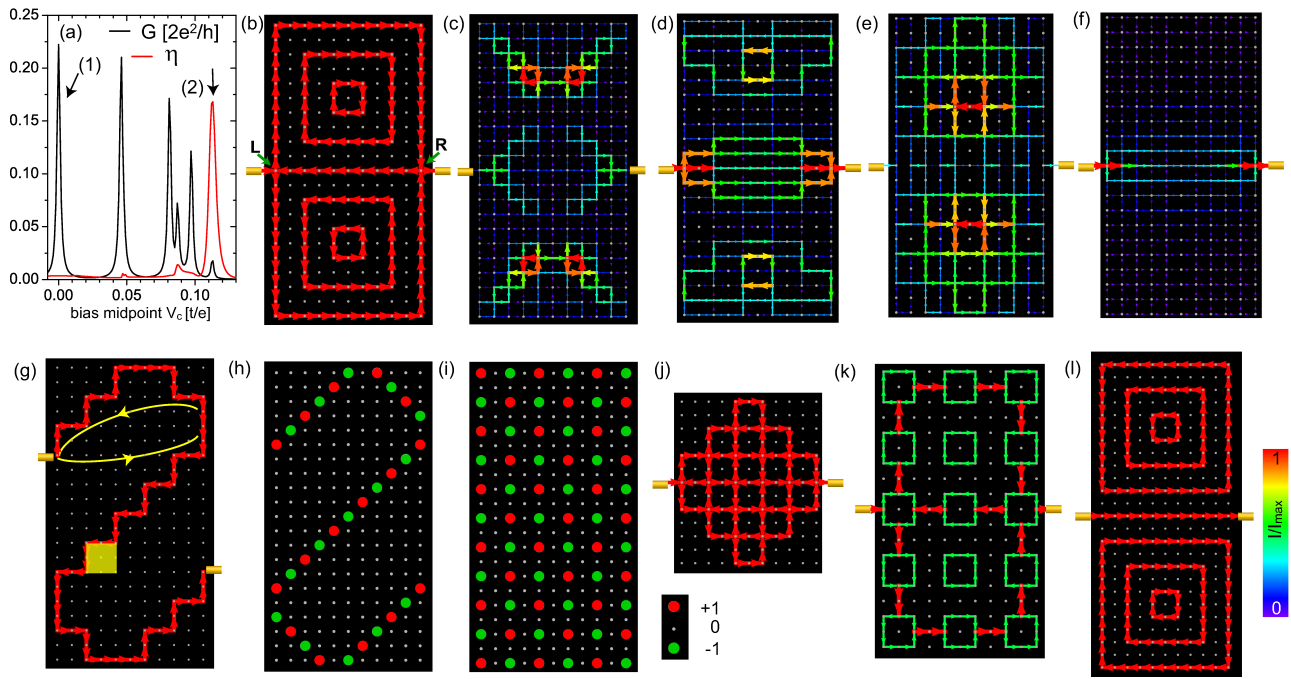} \caption{(a) Conductance, $G$ and the enstrophy, $\eta$.
Spatial current patterns for $\Delta \mu = 0.001t$, $\delta=10^{-3} t$, and (b),(g) $\mu_c = 0$, (c) $\mu_c = 0.113 t $, (d) $\mu_c = 1.819 t $, (e) $\mu_c = 1.932 t $, (f) $\mu_c = 3.912 t $. (h) Real and (i) imaginary part of $G^r_{\bf r,L}$ at $\omega = 0$, normalized by their maximum value. Spatial current patterns for $\mu_c = 0$ in a (j) $(11 \times 13)$, (k) $(11 \times 19)$, and (l) $(11 \times 25)$ network. Currents are normalized by $I_{max}=max|I_{\bf r,r^\prime}|$.} \label{fig:Fig1}
\end{center}
\end{figure*}
For a non-interacting network ($g=0$), one has
$ \hat{G}^{K} =
\left(1-\hat{g}^{r}\hat{t}\right)^{-1}\hat{g}^{K}\left(1-\hat{t}\hat{g}^{a}\right)^{-1}$
with $
{\hat g}^{K}(\omega) = 2i \left[1-2 {\hat n_F}(\omega) \right] {\rm Im}
\left[ {\hat g}^r(\omega) \right]$. Here, $\hat{g}^{r,a,K}$ are the retarded,
advanced and Keldysh Green's function matrices, respectively, containing the Green's functions of the
decoupled ($t,t_h=0$) network sites and leads. This formulation of the problem is sufficiently general such that the network sites can represent atoms, molecules or quantum dots, the
only difference being the form of $\hat{g}$. In what follows, we will assume for simplicity that each site contains only a single, relevant electronic level, such that $g^{r}=1/(\omega +i\delta)$ with $\delta=0^+$, and
will compute the lead Greens functions using the renormalization procedure of Ref.~\cite{Gro89}. Finally, ${\hat n_F}$ is a diagonal matrix containing the
Fermi-distribution functions (with $k_{B} T = 10^{-5} t$ unless otherwise stated) and $\hat{t}$ is the symmetric hopping matrix.

We begin by studying charge transport through the $(11 \times 21)$ non-interacting quantum network ($g=0$) shown in Fig.~\ref{fig:Fig1}. We consider the ballistic (quantum) limit, where
$I_{\bf r  r^\prime}$ does not depend on the chemical potential in the network, and current conservation is automatically satisfied \cite{Car71a}. When the network is disconnected from the
leads, it possesses discrete states with energies  $E({{\bf n}}) =  -2t \left[ \cos(k_{n_{y}}a_{0})+ \cos(k_{n_{y}}a_{0}) \right] $
where $k_{n_{i}}=n_{i}\pi/(N_{i}+1)a_0$, $n_{i}=1,\cdots,N_{i}$ ($i=x,y$), and $a_0$
is the lattice spacing \cite{Tod99}. To identify how these states contribute to the charge transport when two leads are connected to single network sites labeled ${\bf L}$ and ${\bf R}$ [see Fig.~\ref{fig:Fig1}(b)] ,
we plot in Fig.\ref{fig:Fig1}(a) the conductance of the network, $G(V_c) = I(V_{c})/\Delta V$ in the limit of vanishing bias
$\Delta V =(\mu_L-\mu_R)/e \rightarrow 0$, as a function of bias midpoint $V_c=\mu_c/e$ where $ \mu_c=(\mu_L+\mu_R)/2$. In this limit, one obtains $
G(V_c) = 4\pi \frac{e^{2}}{\hbar} t_{h}^4  N_0^2 |G^r_{{\bf L,R}}(\mu_c)|^2
$, where $N_{0}$ is the leads' local density of states, and $G^r_{{\bf L,R}}$ is the non-local Green's function  between ${\bf L}$ and ${\bf R}$ \cite{Car71a}. $G(V_c)$ exhibits a resonance (i.e., a peak) whenever $\mu_c$ coincides with the energy of a network's eigenstate, $E({{\bf n}})$, whose wave function does not vanish at ${\bf L}$ and ${\bf R}$. We refer to these resonances of the conductance as the {\it current eigenmodes} of the network. They are the non-equilibrium analog of the eigenmodes in the (equilibrium) density of states \cite{Man00}, with both types of eigenmodes possessing distinct spatial patterns \cite{com1}.  Note that for the lead positions shown in Fig.~\ref{fig:Fig1}(b), only 121 states (those with odd $n_y$) out of the network's 231 states can carry a current.

Each of the current eigenmodes exhibits a distinct spatial current pattern as shown in Figs.~\ref{fig:Fig1}(b)-(f) [the current patterns shown in Figs.~\ref{fig:Fig1}(b) and (c) correspond to modes (1) and (2) in Fig.~\ref{fig:Fig1}(a), respectively]. Hence, by gating the network (i.e., varying $\mu_c$ ) one can not only alter the total current flowing through the network, but also the spatial path that it takes \cite{Tod99,CLoops,Mar07}.
The current patterns in Figs.~\ref{fig:Fig1}(b) - (e) exhibit loops of circulating currents \cite{Tod99,CLoops},
which are a characteristic feature of the quantum (ballistic) limit, and are absent in classical networks (see below). Note that the current loops in Fig.~\ref{fig:Fig1}(b) are detached from the main current path, and therefore do not contribute to the total current through the network. These loops might be experimentally detectable through the magnetic dipole fields they generate inside the network, in particular, when, as is the case in Fig.~\ref{fig:Fig1}(b), they create a dipole field that is opposite to the one generated by the total current. Associated with the current
loops are links in the network in which the current flows opposite to the applied bias [see center row of Fig.~\ref{fig:Fig1}(b)], a phenomenon referred to as {\it current backflow} \cite{backflow}.
Note that the current pattern for a finite applied bias $\Delta V$ is a superposition of all current patterns associated with current eigenmodes lying between $\mu_L$ and $\mu_R$, as follows directly from Eq.(\ref{eq:Current}).
\begin{figure}
\includegraphics[width=8cm,clip]
{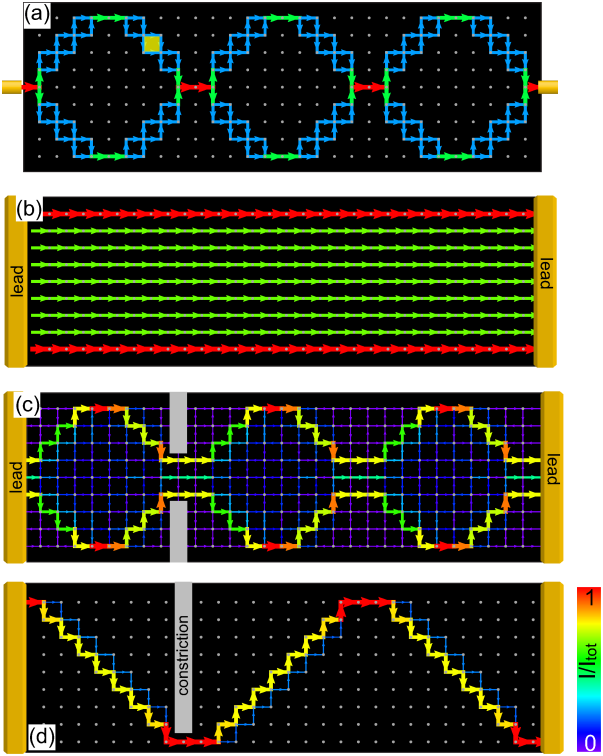} \caption{Current patterns for $\delta = 10^{-4} t$, $\mu_{c} = 0$, $\Delta \mu = 2\times  10^{-3} t$ (a) for narrow leads, and (b),(c) wide leads with constrictions. } \label{fig:Fig2}
\end{figure}

%
%
\begin{figure}[b]
\includegraphics[width=8.5cm]{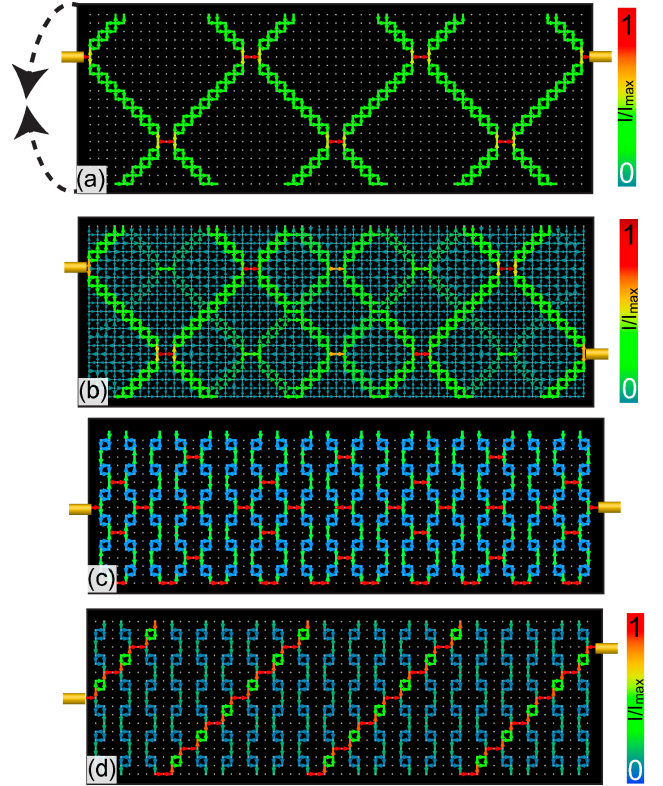} \caption{Current patterns on a cylinder (the cylinder is cut along the axis and flattened out to show the currents) with $\mu_c=0$, $\Delta \mu = 10^{-3} t$, and $\delta=10^{-3}t$. } \label{fig:Fig3}
\end{figure}
In order to understand the spatial structure of these current patterns, we note that for $T=0$ and $\Delta V \rightarrow 0$, the current between two neighboring sites, ${\bf r, r^\prime}$ is
\begin{equation}
I_{\bf r r^\prime}=4 N_{0} t_h^2 \frac{t e^{2}}{\hbar} {\rm Im} \left.
\left[G^{r}_{{\bf r L}} G^{a}_{{\bf L r^\prime}}\right]\right|_{\omega=\mu_c}
\Delta V \label{eq:Ianalytic}
\end{equation}
The contribution from hopping to the right lead is included in the above expression since we made use of the
identity ${\rm Im} \left[G^{r}_{{\bf r L}}G^{a}_{{\bf L r^\prime}} \right] = -{\rm Im} \left[
G^{r}_{{\bf r R}} G^{a}_{{\bf R r^\prime}}\right]$. This form of $I_{\bf r r^\prime}$ demonstrates the non-local character of charge transport in a quantum
network: the current between sites ${\bf r}$ and ${\bf r}^\prime$ arises from electrons that
first propagate from ${\bf r}$ to the left lead and then to site ${\bf r}^\prime$  [as indicated by the yellow arrows in Fig.~\ref{fig:Fig1}(g)], rather than from electrons directly hopping between
the two sites. Eq.(\ref{eq:Ianalytic}) also reveals the microscopic origin of the current's spatial form in more detail. In particular, a comparison of the current pattern shown in Fig.~\ref{fig:Fig1}(g) with
the corresponding real and imaginary parts of the non-local Green's function, $G^{r}_{\bf r L}$, plotted in Figs.~\ref{fig:Fig1}(h) and (i), respectively,  demonstrates that the current path is primarily determined by ${\rm Re} \ G^{r}_{\bf r L}$. Therefore, the current pattern of a given eigenmode does not reflect the spatial structure of the wave-function of the associated eigenstate, which is reflected in the spatial form of ${\rm Im} \ G^{r}_{\bf r L}$, in contrast to earlier findings \cite{Tod99,Mar07}. As a result, the spatial forms of the network's local density of states, $N({\bf r},\omega)=-{\rm Im}G^{r}_{{\bf r,r},\omega}/\pi$ and of the current eigenmode at $\omega = \mu_c$ are in general not related [note that in the limit $t_h \rightarrow 0$, one has $|{\rm Im}G^{r}_{\bf r,r}|=|{\rm Im}G^{r}_{\bf r,L}|$ for $\omega=0$, and therefore in this limit $N({\bf r},\omega=0)$ possesses the same spatial form as ${\rm Im} \ G^{r}_{\bf r L}(\omega=0$  shown in Fig.~\ref{fig:Fig1}(i)]. This conclusion holds in general for all networks and eigenmodes we have considered so far. It is interesting to note that the eigenstate contributing to the current flow for $\mu_c=0$ [Fig.~\ref{fig:Fig1}(b) and (g)] possesses the wave-vector $(k_{n_x},k_{n_y})=(\pi/2,\pi/2)$, and one might therefore expect that the current propagates (on average) along the diagonal direction. However, such a current pattern is only realized if the two leads can be connected by a ``diagonal" path, which is possible for the case in Fig.~\ref{fig:Fig1}(g) [where the path resembles the ballistic propagation of an electron that bounces off the network boundaries], but not for the case in Fig.~\ref{fig:Fig1}(b). Finally, we note that the spatial current patterns exhibit a very rich behavior as a function of the network's aspect ratio $N_y/N_x$, as shown in Figs.~\ref{fig:Fig1}(b), (j)-(l) for fixed $N_x=11$, and varying $N_y$ at $\mu_c=0$.

To determine the effect of current loops on the total current carried by a mode, we consider the normalized enstrophy
$ \eta(V_c)= a_0^2
\ \sum_{\bf r} |\nabla \times {\bf I}_{\bf r}(V_c) |^2/I^{2}_{tot} .
$
Here, the sum runs over all plaquettes of the network, and $\nabla \times {\bf I}_{\bf r}$ is the curl of the current around a plaquette centered at ${\bf r}$. $\eta$ is a measure for the vorticity of the current pattern and thus the extent of the circulating current loops. A comparison of $\eta$ and $G$ [see Fig.~\ref{fig:Fig1}(a), for clarity, only a small range of $V_c$ is shown] reveals that a current mode with a large enstrophy in general carries a small total current. In cases where there are more than one state located between $\mu_L$ and $\mu_R$, one needs to consider the enstrophy and total current carried by each of these states.
Note that for the large enstrophy mode shown in Fig.~\ref{fig:Fig1}(c), the currents inside the network are significantly larger than the total current.

To understand the effect of the leads' width on the charge transport, we consider the $(29 \times 9)$ network shown in Fig.~\ref{fig:Fig2} at $\mu_c=0$. For narrow leads [Fig.~\ref{fig:Fig2}(a)], the current follows a ballistic diagonal path connecting the two leads. In contrast, for wide leads the current flows almost uniformly along the rows of the network, with a larger current flowing along the edge rows [Fig.~\ref{fig:Fig2}(b)].
Such a uniform current pattern can be considered as a superposition of current patterns associated with narrow leads. As a result, one can select narrow lead current patterns [see Fig.~\ref{fig:Fig2}(a)] in wide lead systems by using constrictions as shown in Figs.~\ref{fig:Fig2}(c) and (d), in analogy to mesoscopic systems \cite{Top00}. Moreover, a large degeneracy of a current carrying state exerts a subtle, local effect on the current pattern, as follows from a comparison of Fig.~\ref{fig:Fig1}(g), where the $E=0$ eigenstate is non-degenerate, and Fig.~\ref{fig:Fig2}(a), where the $E=0$ eigenstate is nine-fold degenerate.
While in both cases, the current flows along a diagonal direction, it flows along both sides of a plaquette of width $a_0$ for the $(29 \times 9)$ network [see yellow plaquette in Fig.~\ref{fig:Fig2}(a)]), while for the $(11 \times 21)$ network, it flows only along a single side of a plaquette of width $2a_0$ [see yellow plaquette in Fig.~\ref{fig:Fig1}(g)]). This change in the local current patterns arises from the different spatial dependence of ${\rm Im} \left[G^{r}_{\bf L r} \right]$ in both cases.

To demonstrate that the above qualitative features of current flow  are robust against changes in the topology of a network, we consider in Fig.~\ref{fig:Fig3} a cylinder network (as are realized by carbon nanotubes \cite{Pec03}) with narrow leads at $\mu_c=0$.
For a cylinder with circumference $N_y=20$ and length $N_x=59$  [Fig.~\ref{fig:Fig3}(a)] (where the $E=0$ state is 18-fold degenerate), the two leads are again connected by a diagonal current path that winds around the cylinder; its local and global spatial patterns are similar to those shown in Fig.~\ref{fig:Fig2}(a). When one of the leads is rotated (along the circumference of the cylinder) by an angle of $\pi$ [see Fig.~\ref{fig:Fig3}(b)], it is no longer possible to connect the two leads by a diagonal path, and one obtains a superposition of two current patterns [that shown in Fig.~\ref{fig:Fig3}(a)] which differ by a $\pi$ rotation around the cylinder's axis. In contrast, when the circumference of the cylinder is shortened, as shown in Fig.~\ref{fig:Fig3}(c) for a $(59 \times 18)$ network, the leads cannot be connected by a diagonal path, and the resulting current pattern exhibits a significant amount of transverse current flow, i.e., current flow perpendicular to the direction of the applied bias, as well as current backflow. Only when one lead is rotated by $\pi/3$ along the circumference with respect to the other lead, as shown in Fig.~\ref{fig:Fig3}(d), does the current flow predominantly along a diagonal path connecting the two leads, though some transverse current flow remains.

\begin{figure*} \begin{center}
\includegraphics[height=1.6in,clip]
{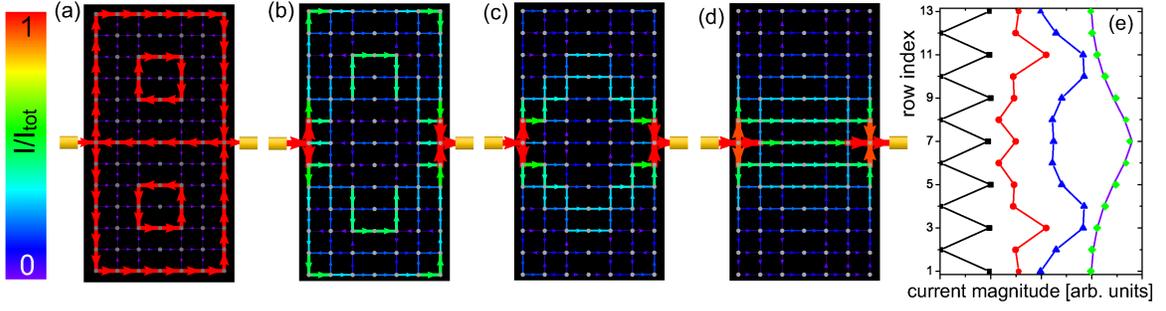} \caption{Spatial current patterns for (a) $\gamma=  0.02t^2$, (b) $\gamma = 0.18t^2$,  (c)  $\gamma = 0.5t^2$, and (d)  $\gamma =8t^2$. Here, $k_{B} T = 100 \omega_{0} = 5\times10^{-5}t$, $\mu_{c} = 0$, $\Delta \mu = 0.06 t$, and leads' Green's function $G^r_l = - i \pi$. (e) Plot of the horizontal current in the middle column of (a) - (d) from left to right and of a classical resistor network (purple line). The current is offset for clarity.} \label{fig:Fig4}
\end{center}
\end{figure*}

In order to study the effects of dephasing \cite{Pec04,Gal07} on the spatial current patterns in nanoscopic networks, we consider the electron-phonon interaction [see Eq.(\ref{eq:Hamiltonian})] \cite{Mon02} and employ the high-temperature approximation $\omega_0 \ll k_B T$ (with $\omega_0 \rightarrow 0$) introduced in Ref.\cite{Bih05}. In this case, the self-energy is related to the full Green's function via $\Sigma^{K,r} = \gamma G^{K,r}$, where $\gamma  = 2g^{2}T/\omega_{0}$, such that the dephasing process is controlled by a single parameter, $\gamma$ (a derivation of the Green's functions in this approximation is presented in Appendix A). The self-consistent solution of the resulting Dyson equations for $G^{K,r}$ guarantees current conservation in the network. In Fig.~\ref{fig:Fig4} we present the evolution of the current pattern for a $(7 \times 13)$ network with increasing $\gamma$. As $\gamma$ increases, the current pattern changes significantly and evolves smoothly from that of the ballistic limit [see Fig.~\ref{fig:Fig4}(a), which is similar to Fig.~\ref{fig:Fig1}(b)] to that of a classical resistor network [see Fig.~\ref{fig:Fig4}(d)]. The evolution is particularly evident when plotting the horizontal current in the middle column [as shown in Fig.~\ref{fig:Fig4}(e)] for the values of $\gamma$ used in Figs.~\ref{fig:Fig4}(a) - (d). For $\gamma = 8t^2$ [corresponding to a dephasing time $\tau \approx \hbar / (4.9t$), see Fig.~\ref{fig:Fig4}(d)], the current pattern is basically indistinguishable from that of the classical resistor network (solid purple line).

In summary, we have shown that nanoscopic quantum networks exhibit current eigenmodes with distinct spatial current patterns. We demonstrated that the rich variety of spatial current patterns arises from the interplay between the network's geometry and electronic structure, the leads' location and width, and the dephasing time, and can be selected via gating or through constrictions. These results suggest new venues for custom-designing current patterns and their transport properties at the nanoscopic, local level. Moreover, we found that a large enstrophy of a current pattern generally corresponds to a small total current carried by the eigenmode.  Finally, we demonstrated that with decreasing dephasing time, the current patterns evolve smoothly from that of a ballistic network to that of a classical resistor network.

We would like to thank P. Guyot-Sionnest, H. Jaeger, and L. Kadanoff for
stimulating discussions. This work is supported by the U.S. Department of Energy
under Award No.~DE-FG02-05ER46225 (D.K.M) and by a Department of Education GAANN Fellowship (T.C.).

\appendix

\section{Appendix A: Dephasing in the presence of phonons}
The Keldysh and retarded Green's function matrices are given by
\begin{align*}
\hat{G}^{K} &= \hat{G}^{r}\left[ \left(\hat{g}^{r} \right)^{-1} \hat{g}^{K} \left( \hat{g}^{a} \right)^{-1} + {\hat \Sigma}^K_{ph} \right] \hat{G}^{a}\\
\hat{G}^{r} &= \hat{g}^{r} + \hat{g}^{r} \left[ \hat{t} + {\hat \Sigma}^r_{ph} \right] \hat{G}^{r}
\end{align*}
where ${\hat \Sigma}_{ph}$ is the fermionic self-energy matrix arising from the electron-phonon interaction. In the limit of temperature being much larger than the phonon frequency $\omega_0$ (i.e., the high-temperature approximation introduced in Ref.~\cite{Bih05}), one retains only those terms that contain a factor of $n_B(\omega_0)$ and the fermionic self-energy at a site ${\bf r}$ in the network in the self-consistent Born approximation is given by
\begin{align*}
\Sigma_{{\bf r r}}^{r,K}(\omega) =  i g^{2} \int \frac{d\nu}{2\pi} D^{K}(\nu) G_{{\bf r r}}^{r,K}( \omega - \nu)
\end{align*}
where
\begin{align*}
D_{0}^{K} = 2 i \pi \left( 1 + 2 n_{B}(\omega)\right) \left[ \delta(\omega + \omega_{0}) - \delta(\omega - \omega_{0})\right]
\end{align*}
is the Keldysh phonon Green's function, which we assume to remain unchanged in the presence of an applied bias, and $n_{B}(\omega)$ is the Bose distribution function. A further simplification is achieved by considering the limit $\omega_{0} \rightarrow 0$ in which the self-energy, to leading order in $T/\omega_{0}$, is given by
\begin{align*}
\Sigma_{{\bf r r}}^{K, r}(\omega) &=  2 g^{2} \frac{T}{\omega_0} G_{{\bf r r}}^{K,r}(\omega) \equiv \gamma G_{{\bf r r}}^{K,r}(\omega)
\end{align*}

We next introduce the superoperator $\tilde{D}$ \cite{Bih05} which, when operating on a Green's function matrix, returns the same matrix with all elements set to zero except for the diagonal elements in the network, e.g.,
\begin{equation}
[\tilde{D} {\hat G}^{r,K}]_{{\bf r r^\prime}} = \left\{
\begin{array}{l}
G^{r,K}_{\bf r r^\prime} \delta_{\bf r,r^\prime} \text{ \ \ \ \ if } {\bf r} \text{ \ \ lies in the network} \\
0\qquad \qquad \text{ \ \ \ otherwise }%
\end{array}%
\right.
\end{equation}
and thus
\begin{equation}
\Sigma^{r,K}(\omega) = \gamma \tilde{D} {\hat G}^{r,K}
\end{equation}
We next define the operator ${\hat U}$ that acts on a matrix ${\hat X}$ via
\begin{equation}
{\hat U}{\hat X} = {\hat G}^{r}{\hat X}{\hat G}^{a}
\end{equation}
The solutions of the above Dyson equations are then given by
\begin{align}
\hat{G}^{K} &= \hat{U} \left[ 1 - \gamma {\tilde D} {\hat U} \right]^{-1} {\hat \Lambda} \\
\hat{G}^{r} &= \left[ 1 - \hat{g}^{r} \left( \hat{t} + \gamma {\tilde D} {\hat G}^r \right) \right]^{-1} \hat{g}^{r}
\end{align}
where we defined the diagonal matrix $\hat{\Lambda} = \hat{g}_{r}^{-1} \hat{g}_{K} \hat{g}_{a}^{-1}$. Note that the only non-zero elements of $\hat{\Lambda}_{\bf r r}$ are those where ${\bf r}$ is a lead site. These elements also contain the chemical potentials of the left and right leads. By expanding the right hand side of the first equation, we obtain
\begin{align}
\hat{G}^{K}_{\bf r r^\prime}  & =  \sum_{\bf l} {\hat G}_{\bf r l}^{r} \left [{\hat \Lambda}_{\bf l l} + \gamma \sum_{\bf m} {\hat Q}_{\bf lm} {\hat \Lambda}_{\bf m m}  \right. \\
& \quad \left. +  \gamma \sum_{\bf m,p} {\hat Q}_{\bf lm} {\hat Q}_{\bf mp} {\hat \Lambda}_{\bf p p} + ... \right] {\hat G}_{\bf l r^\prime}^{a} \nonumber
\end{align}
where
\[
{\hat Q}_{\bf lm}=\left\{
\begin{array}{l}
\left|G_{\bf lm}^{r}\right|^{2} \text{ \ \ \ \ if ${\bf l}$ lies in the network}\\
0 \text{ \ \ \ \ \ \ \ \ \ \ \ \ otherwise}%
\end{array}%
\right.
\]
Defining next the vector ${\bm \lambda}$ with ${\bm \lambda}_{\bf m} = {\hat \Lambda}_{\bf m m}$, we finally obtain
\begin{equation*}
\hat{G}^{K}_{\bf r r^\prime}  =  \sum_{\bf l} {\hat G}_{\bf r l}^{r} \left [ \left(1 - \gamma {\hat Q} \right)^{-1} {\bm \lambda} \right]_{\bf l} {\hat G}_{\bf l r^\prime}^{a}
\end{equation*}
or $\hat{G}^{K} =  {\hat G}^{r} {\tilde \Sigma} {\hat G}^{a}$
where the diagonal matrix ${\tilde \Sigma}$ is defined via
\begin{equation}
{\tilde \Sigma}_{\bf ll}=  \left [ \left(1 - \gamma {\hat Q} \right)^{-1} {\bm \lambda} \right]_{\bf l} .
\end{equation}

\end{document}